\documentclass[a4paper, 10pt, twocolumn]{article}

\usepackage[utf8]{inputenc}         
\usepackage[english]{babel}          

\usepackage{graphicx}               
\usepackage{tabularx}               
\usepackage{multirow}               
\usepackage{url}                

\usepackage[small,bf]{caption2}     
\usepackage{parskip}
\usepackage{titlesec}
\usepackage{amsmath}    
\usepackage{hyperref}
\usepackage{booktabs}
\usepackage{pdflscape}

\titleformat{\section}{\normalfont\large\bfseries}{\thesection}{}{}
\titleformat{\subsection}{\normalfont\large\bfseries}{\thesection}{}{}
\titleformat{\paragraph}{\normalfont\bfseries}{\theparagraph}{}{}
\titlespacing{\section}{0pt}{6pt}{-1pt}
\titlespacing{\subsection}{0pt}{3pt}{-1pt}
\titlespacing{\paragraph}{0pt}{3pt}{-1pt}

\newcolumntype{Y}{>{\centering\arraybackslash}X}    

\usepackage{tabu}

\begin{document}

\date{}                                         

\title{\vspace{-8mm}\textbf{\large
Dataset FAN-01: Revisiting the EAA Benchmark for a low-pressure axial fan }}

\author{
Stefan Schoder$^1$, Felix Czwielong$^2$  \\
$^1$ \emph{\small TU Graz, Aeroakustik und Vibroakustik, IGTE, Inffeldgasse 18, A-8010 Graz, AT, Email: stefan.schoder@tugraz.at
} \\
$^2$ \emph{\small FAU Erlangen-Nürnberg, Aerodynamik und Akustik, LSTM, Cauerstraße 4, 91058 Erlangen, DE
}
} \maketitle
\thispagestyle{empty}           

\section*{Abstract}
We revisited the online repository (\href{https://zenodo.org/communities/eaa-computationalacoustics}{EAA Benchmark datasets for acoustics at Zenodo}), the data source, the details of the experiments, and selected numerical results of the \href{https://zenodo.org/records/8418341}{EAA benchmark case of a low-pressure axial fan in a short duct} \cite{EAABenchmark-FAN}. We present the whole aeroacoustic simulation process and its validation by experimental results. A successful computation of the flow and acoustic involves the following procedure, including validation:
    1. Obtaining a verified mesh for the flow simulation.
    2. Evaluating turbulence modeling.
    3. Performing a mesh convergence study.
    4. Validating significant flow results concerning a subsequent acoustic simulation.
    5. Establishing a computational domain and meshing regarding the acoustic field.
    6. Acoustic source term calculation includes possible truncation and interpolation from the flow grid to the acoustic mesh.
    7. The acoustic field computation and validation of its significant physical quantities.
The data acquisition was motivated to cover all validation steps with experimental data regarding this conceptual workflow. We will present physical and methodological details for a successful aeroacoustic fan simulation.

\section*{Motivation}
Considering arbitrary domains, initial, and boundary conditions, solution techniques (analytic, numerical, and experimental) require certain assumptions and simplifications. Verification and validation of assumptions and results rely on high-quality data for comparison (benchmark). The European Acoustics Association (EAA)-benchmark platform \cite{Hornikx2015,EAABenchmark} provides easily accessible benchmark data for various cases for validating computational acoustics. The Technical Committee of Computational Acoustics of the EAA initiated and manages the platform. We kindly invite you to explore the results of the benchmark cases, validate your calculation, report your results, and encourage you to submit new benchmark contributions over Zenodo by referencing "EAA Computational Acoustics" as community when uploading a dataset valuable for computational acoustics. For more information about the submission process, please see the \href{https://zenodo.org/communities/eaa-computationalacoustics/curation-policy}{EAA Computational Acoustics Data Curation Policy}.

Regarding this overall mission of the EAA-benchmark platform, we will now focus on the specific benchmark of a low-pressure axial fan \cite{Zenger2016}. Sound design and optimization require robust aeroacoustics predictions using numerical methods. Benchmarking computational schemes are of outermost importance and require valid data. Table \ref{tab:summary} summarizes the download link, the available data and benchmark workshops organized similar to the Francis 99 idea \cite{Francis99,Turbine}. A detailed description of the experimental setup and the measurement positions are given in \cite{Zenger2016}.

\begin{table*}[]
\caption{Summary of the benchmark and the available data of the EAA low-pressure axial fan benchmark \label{tab:summary}. \textit{Note that this repository was moved by the EAA form https://www.tuwien.at/en/mwbw/mec/e325-03-research-unit-of-measurement-and-actuator-technology/eaa-benchmarks/benchmarks/acoustics-involving-heterogeneous-and-moving-fluids to Zenodo during the recreation of the repository, when adding the dataset for FSAI \cite{FSAI}.}}
\centering
\begin{tabular}{l p{10cm}}
\toprule
 Hosted by &  EAA \\
 Short Name & Dataset FAN-01\\
 Full Name & Benchmark-Dataset FAN-01: Low pressure Axial Fan in a short Duct\\
 Data: DOI & \url{https://doi.org/10.5281/zenodo.8418341}  \\ 
 Data-Paper: DOI & \url{https://doi.org/10.48550/arXiv.2211.12014}\\
 Citation &  S. Schoder, and Czwielong, F. (2022). Dataset FAN-01: Revisiting the EAA Benchmark for a low-pressure axial fan. arXiv preprint arXiv:2211.12014.\\
 Initial &  F. Zenger, C. Junger, M. Kaltenbacher, and
S. Becker. A benchmark case for aerodynamics
and aeroacoustics of a low pressure axial fan. In
SAE T.P. 2016-01-1805, 2016. \\
 URL &  \url{https://zenodo.org/records/8418341}\\‚ 
 Workshops & 1$^st$ at DAGA 2022 in Stuttgart
 \\
 \toprule
 \textbf{Fan} &    \\
 Design parameters &  See Table \ref{tab:fan_characteristics}  \\
 Geometry &  The rotor geometry is available as IGS-CAD model or Parasolid file.  \\
 Operating condition &  See Table \ref{tab:fan_characteristics}  \\
 \toprule
 \textbf{LDA} &    \\
 LDA system & 2-component LDA probe, type 2D FiberFlow
(Dantec Dynamics)
BSA P80 burst spectrum analyzer (Dantec Dy-
namics)
BSA Flow software v5.20 (Dantec Dynamics)\\
Measurement time & 12 min or $2.5\times10^6$ samples per position \\
\toprule
 \textbf{Wall pressure} & \\
 \textbf{fluctuations} & \\
Type & Differential pressure transducers XCS-093-1psi D
(Kulite Semiconductor Products) \\
Measurement time & 30 s with a sampling frequency of 48 kHz \\ 
\toprule
 \textbf{Microphone} &   \\
 Type & 1/2 inch free-field microphones 4189-L-001 (Brüel
\& Kj\ae r) \\
Measurement time & 30 s with a sampling frequency of 48 kHz \\
\toprule
 \textbf{Microphone array } &  \\
 \textbf{microphones} &  \\
Type & 1/4 inch array microphones 40PH-Sx (G.R.A.S.
Sound \& Vibration) \\
Measurement time & 30 s with a sampling frequency of 48 kHz \\
 \bottomrule
\end{tabular}
\end{table*}

\section*{Benchmark}
\begin{figure}[hbt]
	\begin{center}
		\includegraphics[width=0.9\columnwidth]{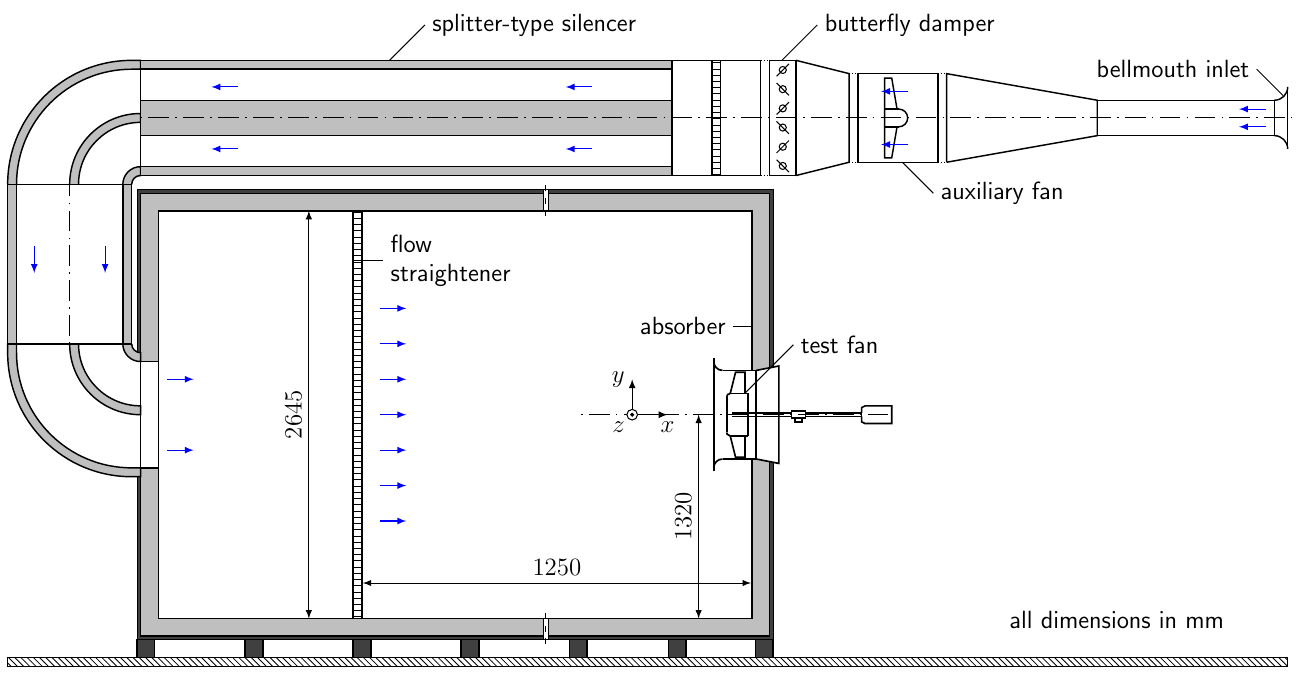}
	\end{center}
	\caption{Standardized inlet test chamber.}
	\label{fig:schema}
\end{figure}

We summarize the available experimental results for comparison with a short review. The extensive amount of measurement data includes aerodynamic performance (volume flow rate, pressure rise, and efficiency), wall pressure fluctuations in the duct (see Fig. \ref{fig:schema}), fluid mechanical quantities (velocity in three spatial directions and turbulent kinetic energy) on the fan suction and pressure side (see Fig. \ref{fig:schema}). More information about the experimental setup can be found in \cite{Zenger2016}.

The fan was designed with the blade element theory for low solidity fans. In terms of size and operating conditions (see Tab. \ref{tab:fan_characteristics}), it is a typical fan to be used in industrial (heating, ventilation, air conditioning - HVAC) and automotive (thermal management cooling fan) applications. The fan has zero blade skew, and the design was not optimized for fluid dynamic or acoustic behavior. The blades consist of NACA 4510 profiles. Table \ref{tab:fan_characteristics} shows the design parameters. The circumferential velocity at the blade tip corresponds to a Mach number of 0.113. Thus, the flow can be considered incompressible. The Reynolds number based on the chord length is almost constant over the span-wise direction of the blade and indicates a turbulent flow. The reproduction of the geometry has been the primary goal. This geometry makes it easy for other scientific groups to develop the benchmark data further using their version of the fan. The data of the original geometry is available online (see Tab. \ref{tab:summary}). All measurements were made in a standardized anechoic inlet test chamber (see Fig. \ref{fig:schema}) according to ISO 5801.

\begin{table}[hb!]
    \caption{Fan design parameters (according to \cite{Zenger2016})}
    \label{tab:fan_characteristics}
    \begin{center}
        \begin{tabular}{ll}
            \toprule
            Fan diameter & $495\,$mm \\
            Hub diameter & $248\,$mm \\
            Tip clearance & $2.5\,$mm \\
            Blades & 9 \\
            Volumetric flow & $1.4\,$m$^3/$s \\
            Total-to-static pressure difference & $150\,$Pa \\
            Rotational speed & $1486\,$1/min \\
            Circumferential velocity hub & 19.4\,m/s \\
            Circumferential velocity tip & 38.9\,m/s \\
            Chord length hub & $103\,$mm \\
            Chord length tip & $58\,$mm \\
            Reynolds number hub & $1.25\cdot 10^5$ \\
            Reynolds number tip & $1.50\cdot 10^5$ \\
            \toprule
        \end{tabular}
    \end{center}
\end{table}

\section*{Data}

The volumetric flow $\dot{V}$ at the operating condition was adjusted by butterfly dampers and an
auxiliary fan in the inlet section. A flow straightener rectified the flow field in the first half
of the inlet chamber. The duct was installed in the
chamber wall, with the suction side facing inwards and the pressure
side facing outwards. An electrical motor
outside of the measurement chamber drove the fan. For more information about measurement setup, we refer to \cite{Zenger2016}. The measured pressure rise of the fan
at the design point is $\Delta p = 126.5\,$Pa. However, the design pressure
difference is not entirely reached due to unconsidered losses like tip flow.
At the design point, the efficiency 
is
$\eta= 53\,$\%.

\subsection*{Geometry}
The CAD geometry is available online at the EAA benchmark platform.

\subsection*{LDA}
\begin{figure}[hbt]
	\begin{center}
		\includegraphics[width=0.98\columnwidth]{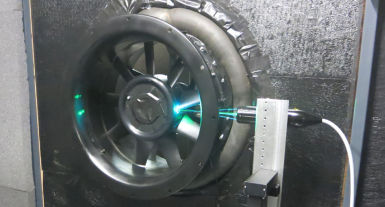}
	\end{center}
	\caption{Picture of the LDA measurement series.}
	\label{fig:lda}
\end{figure}

LDA measurement (see Fig. \ref{fig:lda}) of the flow velocities (meridional, radial, circumferential) on the suction and the pressure side are available. Both ensemble-averaged and time-averaged signals are available. The data can be used to determine the turbulent kinetic energy on the suction and pressure side to validate the turbulent structures inside the inflow.

\subsubsection*{Wall pressure measurements}

\begin{figure}[hbt]
	\begin{center}
		\includegraphics[width=0.9\columnwidth]{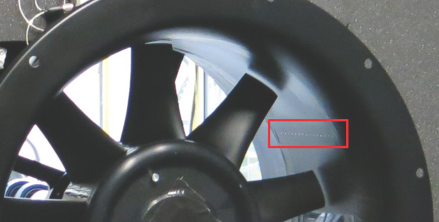}
	\end{center}
	\caption{Picture of the wall pressure sensor distribution inside the duct.}
	\label{fig:Pres}
\end{figure}

The wall pressure fluctuation measurements consist of data from 15 transducers that were installed 15 mm after the end of the nozzle, with a spacing of 10 mm (see Fig.~\ref{fig:Pres}). The pressure probe measurements can validate the effects of the tip gap flow. 

\subsubsection*{Microphone}
\begin{figure}[hbt]
	\begin{center}
		\includegraphics[width=0.9\columnwidth]{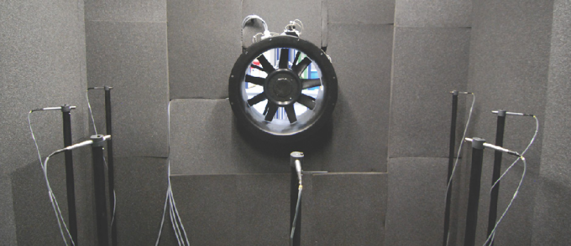}
	\end{center}
	\caption{Picture of the microphone positions inside the anechoic measurement box.}
	\label{fig:Mic}
\end{figure}

Additionally, the acoustic spectra at seven different microphone positions upstream of the fan is provided (see Fig. \ref{fig:Mic}).
The available measurement results contain microphone signals with a 48\,kHz sampling frequency and measurement length of $T=30$\,s. Inside the test chamber, seven ($N=7$) microphones
were installed horizontally in a half-circle with a radius of 1\,m in front of the nozzle
of the duct (see Fig. \ref{fig:Mic}).
\begin{figure}[h!]
  \begin{center}
      \includegraphics[width=0.47\textwidth]{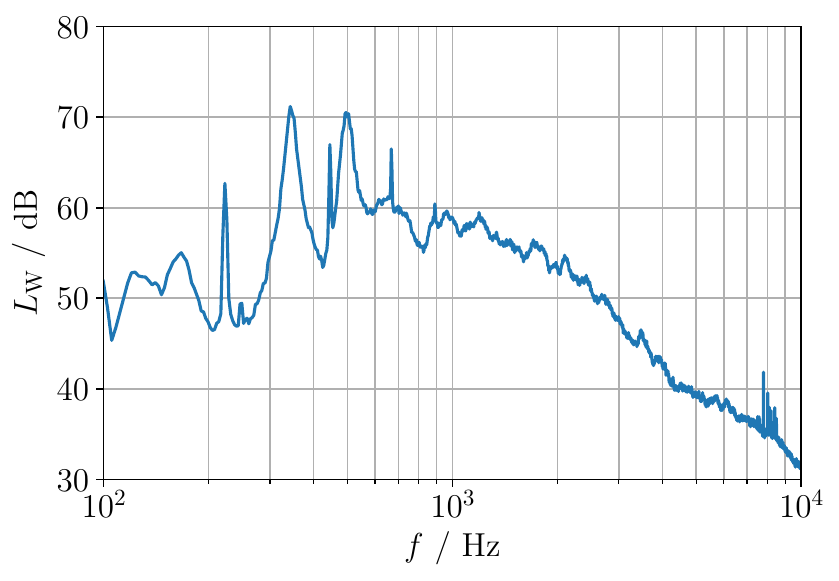}
      \caption{Measured sound power level of the investigated fan.}
      \label{fig:soundpower}
  \end{center}
\end{figure}
The sound power
level was computed by
\begin{equation}
    L_{\rm W} = \bar{L}_{\rm P} + 10 \log{\left(\frac{S_1}{S_0}\right)} ~
    \text{dB}\,,
    \label{eq:soundpowerlevel}
\end{equation}
with the time averaged sound pressure level $\bar{L}_{\rm P}$, the hull of
the measurement area $S_1=6.28\,$m$^2$ and $S_0=1\,$m$^2$. The time averaged
sound power level for all microphones is computed as
\begin{equation}
    \bar{L}_{\rm P} = 10 \log{\left(\frac{1}{N}\sum_{n=1}^{N}
    \frac{1}{T}\int p_n^2
     /p_0^2 \mathrm{d}t \right)} ~ \text{dB}\,,
\end{equation}
with the reference pressure $p_0=20\,\mu$Pa.
For a frequency
range of $100\,$Hz to $10\,$kHz, the measured sound power level was $L_{\rm
W}=87.3\,$dB.
The spectrum of the sound power level is shown in Fig. \ref{fig:soundpower}.
The first Blade Passing Frequency (BPF) can be seen as a sharp peak at
223\,Hz, the first higher harmonic of the BPF at 446\,Hz and the second higher harmonic of the BPF at 675\,Hz. Broader peaks
are around 340\,Hz and 500\,Hz that exceed the ones from the BPF. The first and
the second visible sub-harmonic peak are
expected to result from the interaction of the tip flow with the blades.
 Above 800\,Hz, the spectrum consists of
broadband noise.

\subsubsection*{Microphone Array}

\begin{figure}[hbt]
	\begin{center}
		\includegraphics[width=0.9\columnwidth]{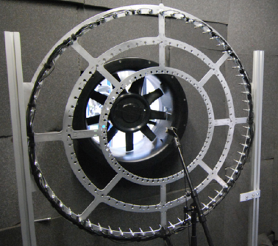}
	\end{center}
	\caption{Picture of the position of the microphone array relative to the axial fan.}
	\label{fig:MicArray}
\end{figure}

Results of the microphone array measurements (see Fig. \ref{fig:MicArray}) are provided online. 

\subsection*{Usage}
The data of the benchmark can be used freely. The community EAA hosting the benchmark encourages to report successful applications to the benchmark organizers.

\section*{Selected validation example}

\begin{figure*}[ht!]
	\begin{center}
		\includegraphics[width=1.0\textwidth]{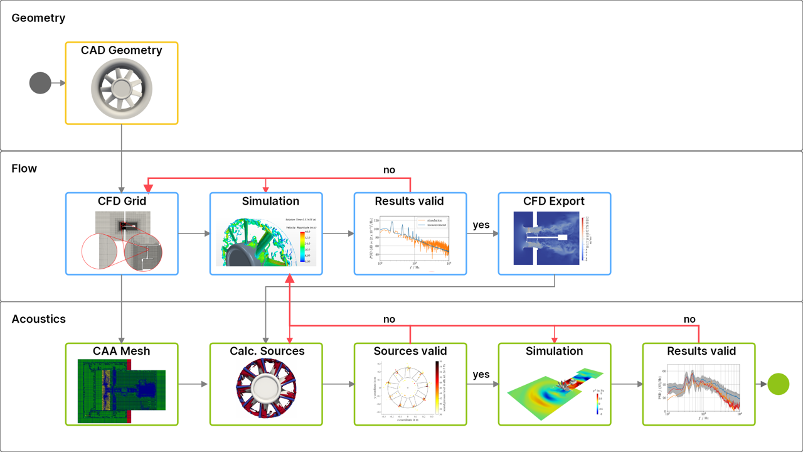}
	\end{center}
	\caption{Validation workflow to perform hybrid aeroacoustic simulations based on a step-by-step validation and a full integration of the benchmark data to validate the whole workflow.}
	\label{fig:Valid}
\end{figure*}

\begin{figure}[hbt]
	\begin{center}
		\includegraphics[width=0.9\columnwidth]{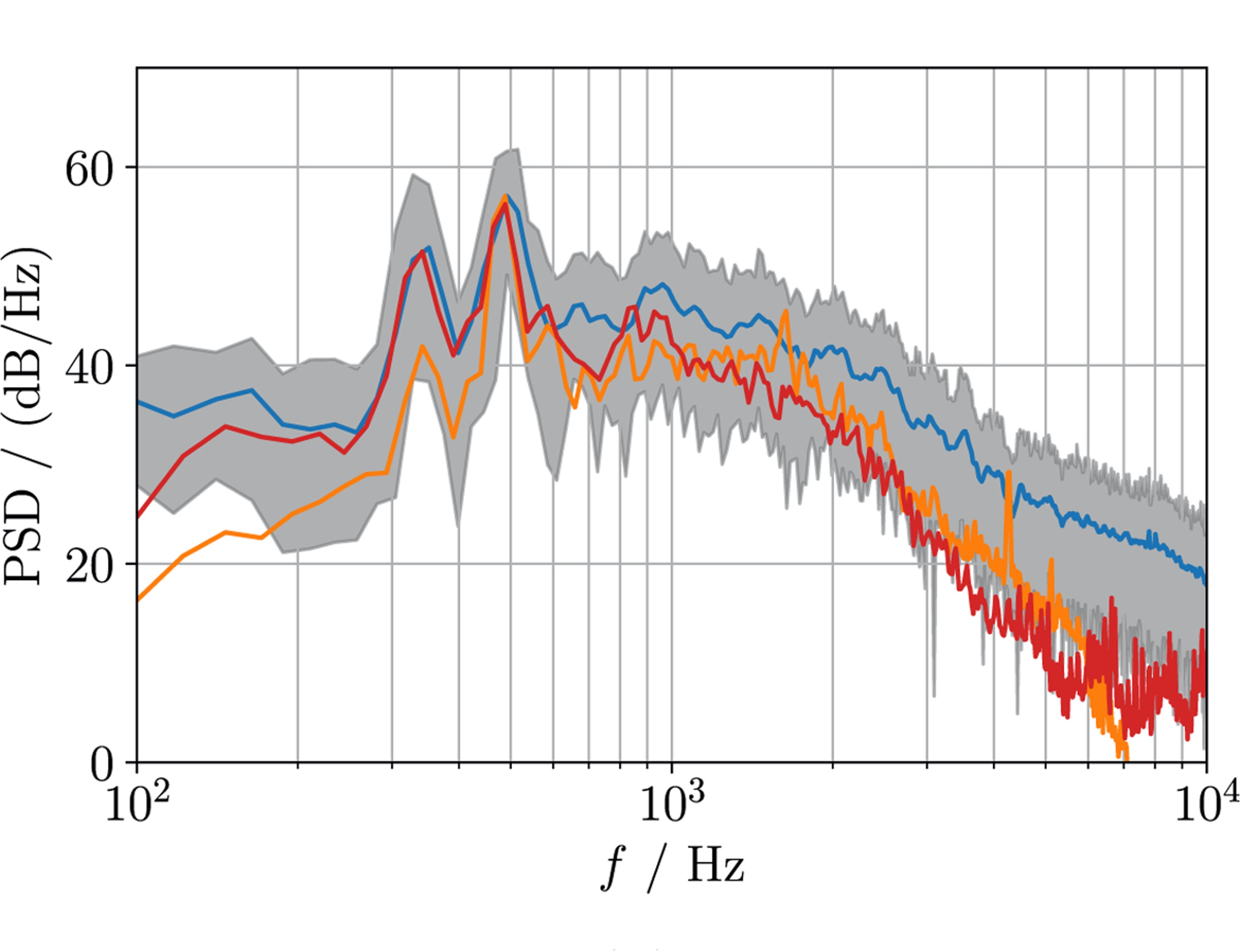}
	\end{center}
	\caption{Results of the aeroacoustic simulations at microphone position 2, with the PCWE result in orange and the FWH result in red. Furthermore, for comparison, measurements over 30 s are depicted in blue, and measurements over 0.1 s are depicted in grey.}
	\label{fig:sim}
\end{figure}

The detailed experimental and numerical results of the EAA benchmark are compared in \cite{Junger2019}. Experimental results validate the whole aeroacoustic simulation process. A successful computation of the flow and acoustic involves the following procedure, including validation. The flow grid is established starting from a valid geometry. Obtaining a verified mesh for the CFD simulation is crucial for the subsequent steps. This verification step includes 
 validation of the turbulence modeling and a flow convergence study based on the grid convergence index. To the end of the flow simulation, it is significant to validate the flow results with global and unsteady local quantities. The unsteady local quantities connected to the aeroacoustic source term of the wave equation are essential and must be traced. This source term is, for instance, the incompressible flow pressure for the perturbed convective wave equation (PCWE). If the flow is valid for the next steps, we proceed with the source calculation and the acoustic simulation.
Now, the sources are computed and transformed while conserving the energy to the acoustic mesh \cite{schoder}. The sources are validated using microphone array measurements. After verifying the sources and the verification of the acoustic mesh, the acoustic computation can be carried out. Final validation of the overall hybrid (partitioned or segregated) workflow \cite{hybrid} can be achieved with the acoustic simulation results. Details on the validation of these steps can be found in \cite{Junger2019} using openCFS \cite{CFS,cfsdata}. Figure  \ref{fig:sim} shows the final validation plot and comparison of the experimental results, the PCWE, and the Ffowcs Williams and Hawkings analogy (FWH). Further investigation on this benchmark were carried out in \cite{Fan1,Fan2,Fan3,Fan4,Fan5,Fan6,Fan7,Fan8,Fan9}. During the investigations of \cite{Fan10,Fan11}, interaction effects of fan noise and duct modes were investigated, which could be of special interest in the future.

\section*{Summary and Benchmark Proposal}

This benchmark is intended to use for validation for numerical methods to predict flow-induced sound. It mainly benefits from the available data that can be used to validate each step of the energy conversion process from the flow to the acoustic field. During the DAGA 2022, several contributions simulated this EAA-Benchmark case and validated their simulation software. For example, \cite{Junger2019} showed aeroacoustic simulations and validated the results of the PCWE and FWH by the benchmark data. Thereby, the authors covered every validation step proposed by the benchmark authors. First, the unsteady wall pressure validates the turbulent flow of the fan. Second, the location of the source terms is compared to microphone array measurements. And finally, a good agreement of the numerical results with the acoustic propagation was presented. In the future, we are curious about your results on this benchmark.

\bibliographystyle{plain}

\end{document}